\documentclass[journal]{vgtc}                     
\usepackage{amsfonts,amssymb,amsmath}
\usepackage{graphics}
\usepackage{epsfig}
\usepackage[all]{xy}
\usepackage{algpseudocode}

\def\Re{{\mathbb{R}}}

\title{Look-and-Twist: \\ A Simple Selection Method for Virtual and Augmented Reality}

\author{%
  Anna Yershova,
   Elmeri Uotila, Katherine J. Mimnaugh, Nicoletta Prencipe,\\
  M. Manivannan, Timo Ojala, and Steven M. LaValle
}

\authorfooter{
  \item
  	All authors except M. Manivannan are with the Center for Ubiquitous Computing, University of Oulu, Finland.
  	E-mail: firstname.lastname@oulu.fi.
  \item
  	M. Manivannan is with the Dept. of Applied Mechanics, Indian Institute of Technology (IIT) Madras, Chennai, India.
  	E-mail: mani@iitm.ac.in.
}

\abstract{%
  This paper introduces a novel interaction method for virtual and augmented reality called {\em look-and-twist}, which is directly analogous to point-and-click operations using a mouse and desktop.  It is based on head rotation alone and is straightforward to implement on any head mounted display that performs rotational tracking.  A user selects features of interest by turning their head to face an object, and then performs a specified rotation along the axis of the looking direction.  The look-and-twist method has been implemented and tested in an educational context, and systematic user studies are underway.  Early evidence indicates that the method is comparable to, or faster than, the standard dwell time method. The method can be used, for example, with Google Cardboard, and it is straightforward to learn for inexperienced users.  Moreover, it has the potential to significantly enrich VR interactions by providing an additional degree of freedom of control, which the binary nature of dwell-based methods lacks.

}

\keywords{}

\renewcommand{\manuscriptnotetxt}{}

\nocopyrightspace

\graphicspath{{figs/}{figures/}{pictures/}{images/}{./}} 

\usepackage{mathptmx}                  

\begin{document}


\firstsection{Introduction}

\maketitle


Virtual, augmented, and extended reality (VR/AR/XR) researchers and practitioners alike know that interaction mechanisms are crucial for the success of their systems.  A wide variety of increasingly sophisticated methods are offered today, including hand gesture recognition, eye-gaze tracking, wearable sensors, game controllers, and brain-computer interfaces.  At the same time, to help VR spread to the masses, there is also a desperate need for simple, effective interaction mechanisms that: 1) work on all VR platforms without requiring special hardware, 2) are easy to implement, 3) are very easy for inexperienced users to learn, 4) are efficient in terms of speed, accuracy, and reliability, and 5) do not lead to user fatigue or exhaustion from extensive use \cite{Kuber_Rashedi_2023,Boring_Jurmu_Butz_2009}.

\begin{figure}
    \centering
    \includegraphics[width=6cm]{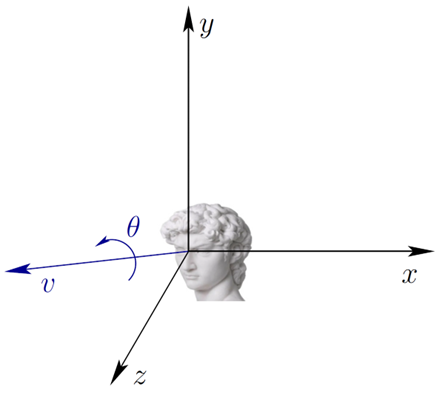}
    \caption{Using Euler's 3D rotation theorem, every 3D rotation can be expressed as a `twist' of $\theta$ about some directional axis $v$ through the origin.  For the look-and-twist method, the user looks in direction $v$ and twists by $\theta$, which both correspond to rotational degrees of freedom, and are easily tracked by a low-cost inertial measurement unit (IMU).}
    \label{fig:euler}
\end{figure}

Pursuing these goals, this paper introduces a simple interaction method called {\em look-and-twist} that is based purely on head rotations, which can be reliably tracked using a standard inertial measurement unit \cite{LavYerKatAnt14,MahHamPfi08}, an already necessary component of any VR system.  The key idea is to carefully separate and utilize the three degrees of freedom associated with rotation.  Utilizing Euler's 3D rotation theorem, any 3D rotation can be expressed using an axis-angle representation:  A rotation by some fixed amount $\theta$, about some rotation axis $v$ that passes through the origin of $\Re^3$; see Figure \ref{fig:euler}.  Suppose the axis corresponds to the direction that the head is facing, and the rotation about that axis is a "twist".  To convert this into an interaction mechanism, the user simply has to {\em look} at a feature of interest, and perform a brief {\em twist} to select it.  In a PC windowing system, these two components are roughly comparable to the mouse operations of moving a pointer to a feature (corresponding to the look part) and clicking on it (corresponding to the twist part).  Thus, the look-and-twist method can be used to select items in the field of view, in a manner similar to point-and-click.

The work was motivated by the desire to make virtual reality as widely accessible to the public as possible.  In the fall of 2024, the authors introduced and taught a course, "Mastering VR: Fundamentals to Practice", at the Indian Institute of Technology Madras, in Chennai, India, with support from the Indian National Programme on Technology Enhanced Learning (NPTEL) and the University of Oulu, Finland.  To make VR affordable to most students across India, the VR platform involved repurposing their smartphones by placing them into low-cost plastic or cardboard viewing cases with lenses. This is much like the highly visible and discontinued Google Cardboard project (2014-2021), but with modern equivalents and carefully customized calibrations to optimize for interpupillary distance, field of view, and lens distortions.  One well-known frustration with these platforms is the lack of interaction mechanisms because the phone's touch screen is inaccessible while placed into the viewer.  Rather than purchase and configure an extra controller, efficient interaction was easy to achieve for hundreds of students by our newly proposed look-and-twist method.  The method is also straightforward to implement on standard platforms such as Unity, Unreal Engine, and Qt Quick 3D XR.  Pointers to implementation details and an online tutorial are provided in Section \ref{sec:method}.  


\section{Related Work} 


Multimodal and hands-free techniques for interacting in virtual environments have been studied comprehensively. Indeed, a 2021 review on hands-free interaction mechanisms discussed 80 papers on the topic \cite{Monteiro_Goncalves_Coelho_Melo_Bessa_2021}, and a more recent review on methods for immersive interaction involving just eye tracking and gaze interactions uncovered over 800 related papers \cite{Plopski_Hirzle_Norouzi_Qian_Bruder_Langlotz_2023}. Eye tracking or head pointing enable the use of intuitive and efficient interaction mechanisms since users look at an object they want to interact with naturally. However, there must be consideration in the design of these interaction techniques because of the "Midas Touch" problem, wherein anything the user looks at triggers a response even though the user does not desire to interact with everything they see as they scan the environment \cite{Jacob_1991}. 

Though there are many possible interaction techniques using gaze which have been investigated, one of the most common methods, called dwell time, involves pointing the head or eyes towards a virtual object and waiting for a specified period to complete the selection. Dwell time can take longer than other methods, but several studies have found higher accuracy rates when using dwell time compared to other techniques \cite{Esteves_Shin_Oakley_2020,Mutasim_Batmaz_Stuerzlinger_2021,Rajanna_Hansen_2018}, though one study did not \cite{Wang_Hu_Chen_2021}. This higher accuracy rate may be due to the ability of the user to correct their choice before the time to confirm the selection has elapsed, whereas if they use some other technique like a button press, they may complete the selection before realizing a mistake has been made \cite{Rajanna_Hansen_2018}.

Dwell time interaction may lead to better accuracy; however, the requirement for a certain amount of time to elapse before completion has led to the creation of alternative methods for selection which can be executed more quickly. For example, Istance et al. \cite{Istance_Hyrskykari_Immonen_Mansikkamaa_Vickers_2010} developed gaze gestures where specific patterns of eye movements generate interactions. Mohan et al. \cite{Mohan_Goh_Fu_Yeung_2018} proposed a method where a confirmation flag pops up next to an item when the user's gaze passes over it, and the user must then look at the flag to trigger the selection. Building on that work, Orlosky et al. \cite{Orlosky_Liu_Sakamoto_Sidenmark_Mansour_2024} created an interaction technique where a small copy of the object (a shadow) becomes visible in peripheral vision when the item is viewed. The user then looks at the shadow for simple selection of the item, or they look at the shadow and back at the object to confirm a manipulation of controllable elements in the environment.

Although methods that utilize eye tracking have been extensively explored, there has been much less work on head pointing as a selection mechanism, and even less work explicitly tackling interaction in mobile VR systems. One study using a Google mobile phone and Daydream VR headset compared a Bluetooth keyboard, dwell time (for 400 ms), a button on the phone, gesture via LeapMotion, motion matching, and speech to make a selection \cite{Esteves_Shin_Oakley_2020}. The results showed that participants preferred using the keyboard and dwell methods, and that participants made fewer errors with dwell and speech compared to the keyboard. In a study using a Samsung smartphone and Gear VR \cite{Xu_Liang_Zhao_Zhang_Yu_Monteiro_2019}, the researchers compared five methods for text entry on a circular keyboard with some swipe motions and dwell time. They found participants made fewer errors and had faster text entry speeds with their dwell-free text entry mechanism. 

Using head tilt as an interaction technique has been investigated previously, although in the context of making a selection which was confirmed by another mechanism, or as an interaction technique mapped to a specific command. In a 2009 study by Crossan and colleagues \cite{Crossan_McGill_Brewster_Murray-Smith_2009}, participants wore a hat with an accelerometer attached to the brim which could move a cursor on a mobile phone, and then they pressed a button to complete the selection. Tregillus, Al Zayer, and Folmer \cite{Tregillus_Al_Zayer_Folmer_2017} utilized head tilt to indicate the direction of motion during a virtual navigation task in a Samsung smartphone and Gear VR. In a 2018 study using a Hololens, Yan et al. \cite{Yan_Yu_Yi_Shi_2018} investigated several head motion gestures for interaction, and used a head tilt to indicate that the user wanted to return to the main menu. The closest work to our proposed method was investigated by Shi and colleagues \cite{Shi_Zhu_Liang_Zhao_2021}, who evaluated task performance and user experience with head tilt and forward-backward head motions for mode-switching tasks in VR. In comparison to these previous methods, the look-and-twist method uses the tilt to confirm a selection or initiate an interaction, which makes the technique fast, flexible, and less prone to errors.

\section{Look-And-Twist Interaction Method}\label{sec:method}

\subsection{Description}

The method is based on a simple idea. Humans have intuition about the direction their head is pointing (the look axis), and it is also quite intuitive to rotate their head around that axis (twist). The resulting look-and-twist orientation of the head represents a head configuration that triggers an interaction with an objects at which the look axis is pointing. This is similar to the dwell time technique in that the object to be selected must first be fixated upon with head gaze. The difference here is that, instead of waiting for a set amount of time to elapse, the head is instead tilted to confirm the selection. 

A demo video of interaction using the look at twist method is available online at: \url{https://youtu.be/X8vq1Nbi8Mg}. A tutorial on implementing the look-and-twist method in Unity (\url{https://youtu.be/ze-0vwsuGko}) and other information about the method (\url{https://lovelace.oulu.fi/iitm-finland/iitm-finland/iitm-hw-twisters/}) are available from the Mastering VR course at IIT Madras.

\subsection{Benefits}

\subsubsection*{Continuous vs Binary Controls.}
The look-and-twist interaction method can be implemented in several different ways. Here, the method was a binary control where a twist triggered a selection regardless of the direction that the head was tilted. However, the method could be implemented with left and right distinct actions. Look-and-twist could also be used as a continuous control, in which the amount of twist can be increased or decreased continuously until a desired level is achieved. The selection can be then finalized by either looking away, or by implementing a dwell timeout.

\subsubsection*{Minimal Hardware Requirements.}

This method can be implemented with any VR system that tracks the user's head rotation. This includes Cardboard VR, which works with the vast majority of current smartphones. 
Although there are various interaction methods that can be considered faster, more efficient, or more comfortable, the significant advantage of the look-and-twist method is that 6 degree of freedom (DOF) tracking, eye tracking, hand tracking, cameras, or controllers are not needed.

\subsubsection*{Easy to Implement.}

The look-and-twist method has the benefit of being very quick and easy to implement.  The implementation requires just a few lines of code to check the current look-at vector and a roll amount around that vector. 
Below is pseudocode of the look-and-twist core:

\begin{algorithmic}
\State $Object \gets VRCamera.RaycastForward()$
\If {$Object \And abs(VRCamera.EulerRotation.z) \geq AngleThreshold$} 
    \State $Object.Interact()$
\EndIf 
\end{algorithmic}

On top of this implementation, a mechanism to prevent repeated interactions is recommended.

\subsubsection*{Ease of Use.}

\begin{figure}
    \centering
    \includegraphics[width=0.5\linewidth]{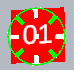}
    \caption{Look-and-twist interaction indicator in neutral position.}
    \label{fig:twistIndicator}
    \vspace{-0.3cm}
\end{figure}

A good interaction mechanism should provide feedback on your progress. For dwell time methods, this is typically represented by a circle that fills up as the dwell progresses, indicating how much time is left before the interaction is completed. In the case of the look-and-twist method, the progress is measured in rotation amount rather than time. Although a similar progress indicator could be used, it may feel more intuitive to have a visual indicator that better matches the motion. We implemented two overlapping circular crosshairs with lines every 90 degrees, in which one circle is controlled by the roll of the camera in VR and the other is in a fixed orientation, with the goal to line the two crosshairs up. This design can be seen in  Figure \ref{fig:twistIndicator}. One challenge here is that the visuals are at a fixed angle; thus, the motion of the crosshair needs to be scaled if the required rotation amount is different than 45 degrees. However, if the required angle is small, there would be less space available for designing the visual indicator of the amount of tilt of the user's head. This visual feedback indicator could be modified to accommodate other designs, for example, by providing a progress bar that fills as tilt occurs. Further research on better visual indicators for this interaction method is merited.

\section{Preliminary User Study}\label{sec:study}

We developed a simple interactive VR task and conducted a quick ad-hoc user study on members in our research lab as preliminary work to evaluate the look-and-twist method in comparison to a dwell time technique.

\subsection{Implementation}

\begin{figure}
    \centering
    \includegraphics[width=0.5\linewidth]{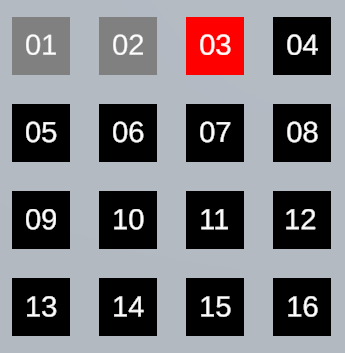}
    \caption{The grid of buttons for the selection task, in which the red button is the next one to be selected.}
    \label{fig:grid}
    \vspace{-0.3cm}
\end{figure}

We implemented a simple menu laid out in a 4-by-4 grid in front of the user that could be interacted with using either the dwell time or look-and-twist method, similar to \cite{Wang_Hu_Chen_2021} (see Figure~\ref{fig:grid}). The application was made in Unity using the OpenXR package for tracking and rendering, but custom-made scripts for the implementation of the interaction methods. Meta Quest 2 was used for this implementation, but positional tracking was disabled to make the tracking functionality match Cardboard VR.

The dwell time interaction simply detected if a button was in the middle of the user's view and triggered the interaction if the specified amount of time was spent looking at the same button. Based on previous research \cite{Wang_Hu_Chen_2021}, we selected a dwell time of 780 ms. If the user looked away or at another button, the timer was reset. After half of the dwell time had passed, a circular progress indicator appeared and filled during the remaining time, with the interaction occurring when the circle was completed. Once the interaction was triggered, the dwell timer reset and would not advance until the user looked away from the current button and turned to look at a button again.

The look-and-twist interaction detected which button was being looked at the same way, but for interaction, required the user to roll the camera on the z-axis by a specified angle, which was 7.5 degrees in our experiment. After the interaction was triggered, the user had to revert their rotation on that axis to 6/9 ths of the required angle for triggering the interaction before being able to trigger another interaction. This way, accidental interactions could be largely prevented even if the required angle for interaction was small. Since this interaction system simply measured the rotation of the camera in VR, interactions could be triggered by tilting the head before looking at the object to be interacted with, then looking at the object. This twist before looking was allowed here since there were no other interactable objects between buttons that could have accidentally been triggered. When the user was looking at a button, a visual indicator as shown in Figure \ref{fig:twistIndicator} became visible. The white lines on the circle were controlled by head rotation with the movement speed scaled up such that they rotated 45 degrees, lining up with the green lines, by the time the interaction was triggered. 

In our experiment, when a user tilted their head 7.5 degrees, the visual indicator rotated 45 degrees. If the user had just interacted with this method, the green part of the indicator turned red until the user sufficiently reverted their head back towards the neutral position. These modes could be switched between by pressing the space bar. The correct button to interact with was determined by a simple sequence, starting from 1 and going in order up to 16. The correct button to press was highlighted in red, previously pressed buttons were gray, and upcoming buttons to press were black. Data was saved into a file whenever the user interacted with a button. If the correct button was interacted with, the time between the last correct button interaction and this interaction was recorded, along with the number of the button and the interaction method. If any incorrect buttons were interacted with, the number of false positives was incremented and saved to the file.


\subsection{Initial Testing}

An initial test was completed with twelve members of our research lab. 
They were given time to familiarize themselves with both of the interaction methods before going through each trial selecting one of 16 buttons in sequence. During practice, interacting with the buttons had no effect, but the participants could see the progress indicator for the interaction and thus confirm that they were able to do the task. Participants were asked if they were ready to start, and once they were, practice mode was turned off and they were instructed to interact with the first button and then all the subsequent buttons up to 16 as fast as possible. After the first trial with the dwell time interaction, there was time for a break, but our participants were ready to proceed right away. Since the program simply measured the time between interactions, the time for pressing the first button was not included in analyses, but the times for the buttons from 2 to 16 were included. Each individual time for a button included the time to move from the previous button to the next, and for the interaction to complete. Therefore, for the dwell time interaction, the average trial time includes the time moving between numbers as well as the 780 ms to complete the selection. In the case of the look-and-twist interaction, the time for each trial included reverting their head position to make the next interaction possible to be triggered. 

The average time to complete a selection across the 15 number selection trials was calculated for each of the participants. For the dwell time interaction method, the mean completion time across all subjects and all trials was 1.1735 seconds (SD = 0.0833), and the mean completion time for the look-and-twist method was 1.1590 (SD = 0.3095). Eleven of the test subjects completed the look-and-twist method for a second time. The average completion time for the second set was 1.0211 seconds (SD = 0.2690).


\section{Discussion}

The time to complete a selection task was compared between a dwell time (780 ms) interaction technique and the look-and-twist interaction technique. Completion times between these two methods were nearly identical the first time that our participants used the look-and-twist method. However, through a second iteration, participants were able to complete the task 0.14 seconds faster than the first time, demonstrating the benefit of an interaction technique which can become easier and more efficient with practice. 

Although one advantage of the look-and-twist method can be seen in shorter task completion times, there are many additional aspects which provide benefit beyond a button selection task. One disadvantage of dwell timers that the look-and-twist method solves is that the user may want to look at something for some period of time without interacting with it. This could be a common scenario when exploring virtual environments where the areas that can be interacted with are part of the environment rather than distinct user interface buttons. With the look-and-twist method, there is a reduced chance that an interaction may be triggered erroneously. 

Furthermore, for exploring virtual environments, a locomotion system is often desirable, which has so far been found difficult to implement without using a separate controller for a VR system that only has 3 DOF tracking. With the look-and-twist method, the user could look at a point on the floor to go to, and then tilt their head in order to teleport to that location. Given that head tilt is not a motion that users normally make during immersion, the look-and-twist interaction would be unlikely to be triggered on accident and could be utilized at will even at a relatively fast pace. Alternatively, even continuous movement could be implemented, for example by looking at the ground in the direction you want to go and tilting your head, continually moving as long as the head is sufficiently tilted. The look-and-twist method could also be used to grab and hold items. 

Another advantage of this method is how simple and intuitive it is to use. Particularly when using a VR viewer with a mobile phone, no extra controllers are needed. For people who do not play computer games often, or for those who may be using VR for the first time (for example, when giving demos), the look-and-twist method may be quick and easy to learn.

One thing to note is that many commercial mobile phone-based VR applications tend to have dwell timers in the range of 1 to 2 seconds. However, in research, the dwell timers are often less than a second. If the required tasks are limited to specific tasks, the interactions can be optimized to be faster. However, shorter dwell timers are more likely to result in false positives, or cause stress to the users trying to avoid accidental interactions. Similarly with the look-and-twist method, the angle threshold for the interaction affects the balance between false positives and negatives. Depending on the user and the task, a higher or lower angle threshold could be appropriate. 

\section{Conclusion}

The look-and-twist method provides a simple and efficient technique to trigger interaction in a virtual environment using head movement alone. With a few lines of code, look-and-twist can be utilized for a VR system with 3 DOF tracking to preclude the need for a controller, or in a system with 6 DOF tracking to provide interaction which can be used simultaneously with a controller. Additionally, it can significantly enrich VR experiences by offering ways to navigate and interact without relying on any additional tracking devices. In initial testing, look-and-twist interactions originally performed comparably with a 780 ms dwell time selection. However, with the look-and-twist method, task completion time decreased on a second iteration. We have provided pseudocode and links to tutorials here so that this method can be easily implemented for VR applications.

\bibliographystyle{abbrv-doi-hyperref}

\bibliography{biblio}


\appendix 

\end{document}